\documentclass[aps]{revtex4}

\usepackage{amssymb}
\usepackage{amsmath}
\usepackage{graphicx}
\usepackage{epsfig,amsmath}

\begin{document}

\title{Numerical simulation of stochastic motion of vortex loops under
action \\ of random force. Evidence of the thermodynamic
equilibrium. }
\author{Luiza P. Kondaurova and Sergey K. Nemirovskii
\thanks{email address: nemir@itp.nsc.ru}}
\affiliation{email address: nemir@itp.nsc.ru \\
Institute of Thermophysics, Lavrentyev Ave, 1, 630090, Novosibirsk, Russia,\\
Novosibirsk State University, Novosibirsk Russia}
\date{September 26, 2008}

\begin{abstract}
Numerical simulation of stochastic dynamics of vortex filaments
under action of random (Langevin) force is fulfilled. Calculations
are performed on base of the full Biot--Savart law for different
intensities of the Langevin force. A new algorithm, which is based
on consideration of crossing lines, is used for vortex
reconnection procedure. After some transient period the vortex
tangle develops into the stationary state characterizing by the
developed fluctuations of various physical quantities, such as
total length, energy etc. We tested this state to learn whether or
not it the thermodynamic equilibrium is reached. With the use of a
special treatment, so called method of weighted histograms, we
process the distribution energy of the vortex system. The results
obtained demonstrate that the thermodynamical equilibrium state
with the temperature obtained from the fluctuation dissipation
theorem is really reached.\newline PACS-numbers: 67.40.Vs 98.80.Cq
7.37.+q\\
\end{abstract}

\maketitle

\section{Introduction}

Quantized vortices appeared in quantum fluids and other systems play a
fundamental role in the properties of the latter. For that reason they have
been an object of intensive study for many years (for review and
bibliography see e.g. \cite{Don}). The greatest success in investigations of
dynamics of quantized vortices has been achieved in relatively simple cases
such as a vortex array in rotating helium or vortex rings. However these
simple cases are rather exception than a rule. Due to extremely involved
dynamics initially straight lines or rings evolve to form highly entangled
chaotic structure. Thus, the necessity of statistic methods to describe
chaotic vortex loop configurations arises. A most tempting way is to treat
vortices as a kind of excitations and to use thermodynamic methods. One of
first examples of that way was an use the Landau criterium for critical
velocity where vortex energy and momentum were applied to relation having
pure thermodynamic sense. More extended examples would be the famous
Kosterlitz--Thouless theory or its 3D variant intensively being developed
currently \{for review and bibliography see e.g. \cite{Williams}). In the
examples above and in many other it is assumed that chaotic vortex
configuration is in thermal equilibrium and their statistics obeys the Gibbs
distribution. That belief is based on fundamental physical principles and
can be justified in a standard way considering vortex loops as a subsystem
submerged into thermostat and exchanging with energy with the latter.
However numerous experiments on counterflowing HeII, and direct numerical
simulations of vortex line dynamics \cite{Don}, \cite{NF}, \cite{Chorin94}
convincingly demonstrate that this dynamics is essentially nonequilibrium
and possesses all features inherent in turbulent phenomena. Thus a question
arises how a thermal equilibrium is destroyed and what mechanisms are
responsible for that. To answer that question we have firstly to understand
in details how a thermal equilibrium in vortex loop configuration space is
established. A general principle of maximum entropy does not give any
details the dynamic details are absorbed by a temperature definition. It is
well known however that the Gibbs distribution can be alternatively obtained
on the basis of some reduced model like kinetic equations or Fokker--Planck
equation (FPE). That way of course is not of such great generality as a
principle of maximum entropy, but instead it allows to clarify the
mechanisms how the Gibbs distribution established \cite{nemirTMF}.

In the presented paper we report preliminary results of numerical
study on dynamics of vortex tangle under action of random
(Langevin) forcing delta correlated both in space and time. The
data obtained were tested to learn whether or not the stationary
state reached in numerical experiment is the thermodynamical
equilibrium state. With use a special treatment, so called method
of weighted histograms, to process distribution energy of the
vortex system. The results obtained demonstrated that the
thermodynamical equilibrium with the temperature obtained from the
the fluctuation dissipation theorem is really reached.
\section{The Numerical Simulation and Results}
We consider the dynamics of vortex loops in three-dimensional space with no
boundaries. The equation of motion of the vortex line elements is supposed
to be:
\begin{equation}
\dot{\mathbf{s}}=\dot{\mathbf{s}}_{B}+\alpha (\mathbf{s^{\prime }\times (v}%
_{n}\mathbf{-\dot{s}}_{B}\mathbf{))-}\alpha \mathbf{^{\prime }s^{\prime
}\times \lbrack s^{\prime }\times (v}_{n}\mathbf{-\dot{s}}_{B}\mathbf{)]+f(}%
\xi ,t\mathbf{),\quad \quad \quad \quad \quad }  \label{motion eq}
\end{equation}%
where $\dot{\mathbf{s}}_{B}$ is the propagation velocity of the vortex
filament at a point $\mathbf{s}$, defined by Biot-Savart low; $\mathbf{s(\xi
,t)}$ is the radius-vector of the vortex line points; $\mathbf{v}_{n}$ is
the normal velocity of the superfluid helium; $\xi $ is a label parameter,
in this case it the arc length; $\mathbf{s^{\prime }}$ is the derivative wrt
the arc length, $\alpha $, $\alpha ^{\prime }$ are the friction
coefficients, describing interaction of vortex filament with normal
component, and $\mathbf{f(}\xi ,t\mathbf{)}$ is the Langevin force. Further
we will take $\mathbf{v}_{n}=0$ and neglect the term with $\alpha ^{\prime }$%
. The Langevin force is supposed to be a white noise with the following
correlator
\begin{equation}
\left\langle \mathbf{f}_{i}(\xi _{1},t_{1})\mathbf{f}_{j}(\xi
_{2},t_{2})\right\rangle \ =D\delta _{ij}\delta \ (t_{1}-t_{2})\delta \ (\xi
_{1}-\xi _{2}).  \label{correlator}
\end{equation}%
Here $i,j$~are the spatial components; $t_{1},t_{2}$ are the arbitrary time
moments; $\xi _{1},\xi _{2}$ define any points on the vortex line; $D$ is
the intensity of the Langevin's force. Let us consider the probability
distribution functional \cite{nemirTMF} defined as
\begin{equation}
\mathcal{P}(\{\mathbf{s}(\xi )\},t)=\left\langle \delta \left( \mathbf{s}%
(\xi )-\mathbf{s}(\xi ,t)\right) \right\rangle .  \label{pdf}
\end{equation}%
The Fokker-Planck equation for the time evolution of quantity $\mathcal{P}(\{%
\mathbf{s}(\xi )\},t)$ can be derived from equation of motion
(\ref{motion eq}) in standard way (see e.g.
\cite{Zinn-Justin96},\cite{nemirTMF})
\begin{gather}
\frac{\partial \mathcal{P}}{\partial t}+\int d\xi \frac{\delta }{\delta
\mathbf{s}(\xi )}\left[ \dot{\mathbf{s}}_{B}+\alpha \mathbf{s}^{\prime }(\xi
)\times \dot{\mathbf{s}}_{B}\right] \mathcal{P}+  \label{FP2} \\
\int \int d\xi d\xi ^{\prime }\left\langle \mathbf{f(}\xi \mathbf{)f(}\xi
^{\prime })\right\rangle \delta (\xi \mathbf{-}\xi ^{\prime })~\delta
(t_{1}-t_{2})~\delta _{\eta _{1},\eta _{2}}~\frac{\delta }{\delta \mathbf{s}%
(\xi )}\frac{\delta }{\delta \mathbf{s}(\xi ^{\prime })}\mathcal{P}=0.
\notag
\end{gather}%
As it was shown in \cite{nemirTMF} the Fokker-Planck equation (\ref{FP2})
has a stationary solution
\begin{equation}
\mathcal{P}(\{\mathbf{s}(\xi )\},t)=\exp (-H\left\{ \mathbf{s}\right\}
/k_{B}T).  \label{Gibbs}
\end{equation}%
Here $H\left\{ \mathbf{s}\right\} $ is the hamiltonian - the energy of the
vortex system expressed via the whole line configuration
\begin{equation}
H\left\{ \mathbf{s}\right\} =E=\int \frac{{\rho }_{s}\mathbf{v}_{s}^{2}}{2}%
\;d^{3}\mathbf{r}=\frac{\rho _{s}\kappa ^{2}}{8\pi }\int\limits_{0}^{L}\int%
\limits_{0}^{L^{\prime }}\frac{(d\mathbf{s}_{1}d\mathbf{s}_{2})}{|\mathbf{s}%
_{1}-\mathbf{s}_{2}|}.  \label{energy}
\end{equation}%
The \textquotedblleft \thinspace temperature\textquotedblright\
$T$ enterring equation (\ref{FP2}) is determined with the help of
the fluctuation dissipation theorem
\begin{equation}
\left\langle \mathbf{f}_{\eta _{1}}\mathbf{(}\xi _{1},t_{1})~\mathbf{f}%
_{\eta _{2}}\mathbf{(}\xi _{2},t_{2})\right\rangle ~=\frac{k_{B}T\alpha }{%
\rho _{s}\pi (\hbar /m)}~\delta (\xi _{1}\mathbf{-}\xi _{2})~\delta
(t_{1}-t_{2})~\delta _{\eta _{1},\eta _{2}}.  \label{fdt_line}
\end{equation}%
In discrete variant of parametrization of the curve, used in numerical
simulation, the delta function $~\delta (\xi _{1}\mathbf{-}\xi _{2})$ is
changed with $1/a$, where $a$ is the step along the curve. Thus intensity $D$
of pumping external Langevin force is connected to the \textquotedblleft
\thinspace temperature\textquotedblright\ via relation

\begin{equation}
D=2kTa/(\rho _{s}\kappa ).  \label{5}
\end{equation}%
Thus, we demonstated that the set of filament agitating by the random white
noise is driven into thermodynamical equilibrium with the temperature
relating to intensity of the random forcing. The main purpose of the the
present work is to demonstrate it in the direct numerical simulation.

Details of numerical simulation were described in our early publication \cite%
{K04}. The new algorithm for vortex reconnection processes basing on the
consideration of crossing lines is used . We run the calculations with $%
\alpha =0.098$, $D_{1}=9.8\cdot 10^{-5}$ cm$^{2}$/s,
$D_{2}=10.2\cdot 10^{-5} $ cm$^{2}$ /s. In our calculations we
start with an initial vortex configuration of twenty four vortex
rings (see Fig. 1 (a)). The initial condition was chosen to make
the total momentum of the system is equal to zero.
\begin{figure}[h]
\centering
\includegraphics[width= 0.9 \textwidth]{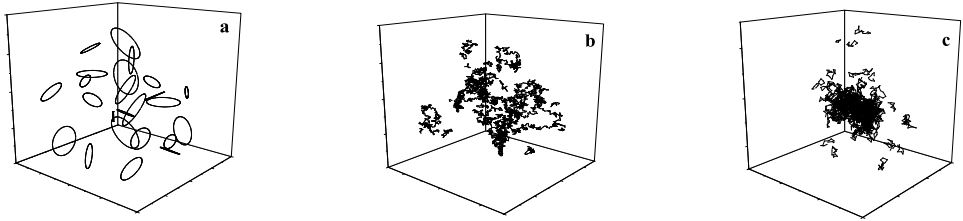}
\caption{ Development of a vortex tangle: t = 0 ($a$) , $t=6$
($b$), $t=120$ ($c$) ms.} \label{fig1}
\end{figure}
To check the behavior of the system we had been monitoring the
total length of the vortex tangle. These quantities for two
different intensities of the random forcing were plotted as
functions of time in Fig. 2. One can see, at the beginning the
total length rapidly increases. When the vortex tangle becomes
dense enough many small vortex loops appear due to reconnection
processes (see Fig.1(b)). These small loops are radiated in the
ambient space and the total length decrease. Finally a stationary
state with the strongly fluctuating is achieved see Fig.1 (c). We
aim now to study this steady state and give some proofs it be in
thermal equilibrium. We use for it the the weighted histogram
analysis method, widely used in the Monte Carlo computer
simulations of the Ising model \cite{F88}.
\begin{figure}[h]
\centering
\includegraphics[width=0.37 \textwidth]{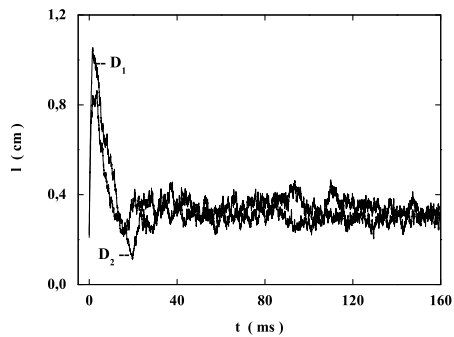}
\caption{ The lengths of lines as a function of time for different
intensity of the Langevin's force $D$.} \label{fig2}
\end{figure}

To apply the weighted histogram analysis method we take for every run $N$
configurations of the vortex tangle at $\ N$ different moments in time. We
supposed these configurations to be statistically independent. We calculate
\ further (With the use of relation (\ref{energy})) the energies of the
every configurations. Dividing then the whole interval of energies in pieces
of width $\Delta E$ we build up the histograms showing relative frequency of
meeting the configuration with the energies lying in the interval between $%
E_{j}$ and $E_{j}+\Delta E$ (see Fig.3). Let us introduce the probability
density $P(E_{j})$ of the observing the vortex system state with the
energies lying in the interval between $E_{j}$ and $E_{j}+\Delta E.$
Obviously, the $P(E_{j})$ can be simply counted \ from histogams with the
following relation
\begin{equation}
P(E_{j})=N(E_{j})/(N\Delta E).  \label{probability hist}
\end{equation}%
Here $N(E_{j})$ is the number of states in the interval between $E_{j}$ and $%
E_{j}+\Delta E$, $N=\sum\limits_{j}N(E_{j})$ is the full number of
configurations (for each of two different intensities of the Langevin's
force $D$ ).
\begin{figure}[h]
\centering
\includegraphics[width=0.37 \textwidth]{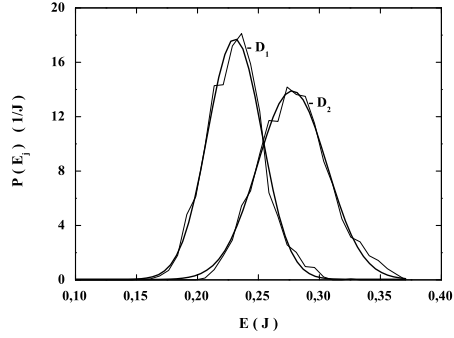}
\caption{ Probability density distributions for different
intensities of the Langevin's force $D$ which are approximated by
Gaussian curves.} \label{fig3}
\end{figure}
Consequently assuming that the vortex loops system is in the thermal
equilibrium we propose that the histograms depicted in Fig.3 can be
describes as well with the use of the Gibbs distribution:%
\begin{equation}
P(E_{j})=\frac{g(E_{j})e^{-E_{j}/k_{B}T}}{\sum%
\limits_{j}g(E_{j})e^{-E_{j}/k_{B}T}},  \label{Gibbs hist}
\end{equation}%
where $g(E_{j})$ is the density of states, and $T$ is the temperature
(either $T_{1}$ or $T_{2}$ for each of the runs), calculated from the
fluctuation dissipation theorem (\ref{fdt_line}),(\ref{5}) for different
intensities of the Langevin's force $D_{1}$, $D_{2}$ correspondingly.
Relation (\ref{Gibbs hist}) included the density of states $g(E_{j})$, which
is not only known but even is not defined well for set of continious curves.
This problem however can be eliminated with the trick offered in paper \cite%
{F88}. Indeed, it is possible to prove that the probability density
distribution at $T_{2}$ can be expressed in terms of the distribution at $%
T_{1}$ in the following way:
\begin{equation}
\tilde{P}_{T_{2}}(E_{j})=\frac{P_{T_{1}}(E_{j})e^{-(\frac{E_{j}}{k_{B}T_{1}}-%
\frac{E_{j}}{k_{B}T_{2}})}}{\Delta E\sum\limits_{j}P_{T_{1}}(E_{j})e^{-(%
\frac{E_{j}}{k_{B}T_{1}}-\frac{E_{j}}{k_{B}T_{2}})}}.
\label{probability transform}
\end{equation}%
Using relation (\ref{probability transform}) we calculated the probability
density $\tilde{P}_{T_{2}}(E_{j})$ for the temperature $T_{2}$ via the
probability density $P_{T_{1}}(E_{j})$ and the compared the result obtained
with the initial histogram for the $P_{T_{2}}(E_{j})$. At this point there
appeared one difficulty. We mentioned that the discrete variant of the
fluctuation dissipation theorem we have to change the delta function $%
~\delta (\xi _{1}\mathbf{-}\xi _{2})$ with quantity $1/a$, where
$a$ is the step along the curve. \ But during evolution the
distance between points does not preserves, the elements of line
either shrink or stretch. We nontheless retain the initial value
of the space step in the definition of the temperature but
introduce the fitting parameter for it. As a fitting parameter we
taken half as much again the maximum step along the vortex line
using in calculation of dynamics of vortex loops. The curves
calculated
according to equations (\ref{probability hist}) \ (for the temperature $%
T_{2} $) and (\ref{probability transform}) are shown in Fig. 4. As it can
seen the curves are very close to each other. It means that the ensemble of
vortex filaments is driven into thermodynamical equilibrium state.
\begin{figure}[h]
\centering
\includegraphics[width=0.37 \textwidth]{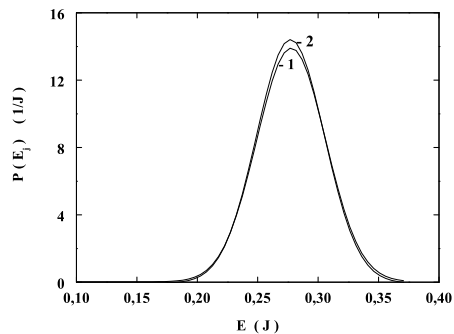}
\caption{ Probability density distributions at the temperature
$T_{2}$: 1 - the curve calculated according to equation (7); 2 -
the curve calculated according to equation (8).} \label{fig4}
\end{figure}

As well it is known that in a thermal equilibrium the variance in the energy
(or `energy fluctuation`) is connected with the temperature in the following
way :
\begin{equation}
\delta E^{2}=Ck_{B}T^{2},
\end{equation}%
were
\begin{equation}
\delta E^{2}=\left\langle (E-\left\langle E\right\rangle )^{2}\right\rangle ,
\end{equation}%
$C=(\left\langle E_{2}\right\rangle -\left\langle E_{1}\right\rangle
)/(T_{2}-T_{1})$ is the heat capacity. We determined energy fluctuations
according to equation (12): $\delta E(T_{1})=2.55\cdot 10^{-16}$ J, $\delta
E(T_{2})=2.65\cdot 10^{-16}$ J, as well as according to equation~ (13): $%
\delta E(T_{1})=2.30\cdot 10^{-16}$ J, $\delta E(T_{2})=2.87\cdot 10^{-16}$ J.
Obtained energy fluctuations are close to each other. It corroborate that
the ensemble of vortex filaments is driven into thermodynamical equilibrium
state too.

\section{Conclusion}

Grounding on results of theoretical study made by one of the authors we
propose that the vortex tangle filaments undergoing the random
\textquotedblleft \thinspace white noise\textquotedblright\ forcing \ is
driven into thermodynamical equilibrium state. The temperature of the vortex
system determined by the fluctuation dissipation theorem. With the direct
numerical simulation we present the proofs of our supposition. The proof was
based on the observation that the distribution of the energy satisfied to
the Gibbs law. It was shown with the method of weighted histograms, widely
used approach in statistical physics.

\section*{ACKNOWLEDGMENTS}

Authors are grateful to S. Chekmarev for numerous discussions and
consultation. This work was partially supported by grants 05-08-01375 and
07-02-01124 from the RFBR and grant of the Russian Federation President on
the state support of leading scientific schools NSH-4366.2008.8.


\begin{thebibliography}{99}
\bibitem{Don} Donnelly, R.J. \textit{Quantized Vortices in Helium II},
Cambridge University Press, 1991.

\bibitem{Williams} G.A.Williams, \textit{Vortex-Loop Phase Transitions in Liquid Helium, Cosmic Strings,
and High-Tc Superconductors}, Phys.Rev.Lett., 1999, \textbf{82}, N6, 1201.
%\TEXTsymbol{\backslash}%

\bibitem{NF} S. K. Nemirovskii and W. Fiszdon,
\textit{Chaotic quantized vortices and hydrodynamic processess superfluid helium},
 Rev. Mod. Phys., 1995, \textbf{67}, N1, 37.
%\TEXTsymbol{%\backslash}textbf\


\bibitem{Chorin94} A. Chorin, \textit{Voticity and Turbulence},
Springer-Verlag, New-Yourk, 1994.

\bibitem{nemirTMF} S.K. Nemirovskii, \textit{Thermodynamic equilibrium in the system of chaotic
quantized vortices in a weakly imperfect Bose gas},
Teoretical and Matematical Physics, 2004,
\textbf{141}, N 1, 141.

\bibitem{Zinn-Justin96} Jean Zinn-Justin, \textit{Quantum Field Theory and
Critical Phenomena}, Claberson Press, Oxford, 1992.

\bibitem{K04} L.P. Kondaurova and S.K. Nemirovskii,
\textit{ Full Biot-Savart numerical simulation of vortices in He II}, J. Low Temperature
Physics, 2005, \textbf{138}, N3/4, 555. %\TEXTsymbol{%\backslash}textbf\

\bibitem{F88} Alan M. Ferrenberg and Robert H. Swendsen,
\textit{New Monte Carlo technique for studying phase transitions},
Phys. Rev. Lett., 1988, \textbf{61}, 2635.

\end{thebibliography}
\end{document}